\documentstyle[epsf,latexsym]{l-aa}
\newcommand{\be}{\begin{equation}}
\newcommand{\ee}{\end{equation}}

\begin{document}
\thesaurus{01 (03.13.2; 11.05.1; 11.09.1; 11.19.5)}
\title{Crowded-field photometry from HST-imaging
\thanks{Based on observations made with the NASA/ESA Hubble Space Telescope, 
        obtained from the data archive at the Space Telescope Science 
        Institute. STScI is operated by the Association of Universities 
        for Research in Astronomy, Inc. under the NASA contract NAS 5-26555.}}
\author{Marianne Sodemann and Bjarne Thomsen}
\offprints{bt@obs.aau.dk}
\institute{Institut for Fysik og Astronomi, Aarhus Universitet, Ny
 Munkegade, DK-8000 \AA rhus C, Denmark}
\date{Received \hskip 2cm; Accepted}
\maketitle

\begin{abstract}
We present a thorough investigation of stellar photometry based
on HST imaging of crowded fields at $85\arcsec$ and $10\arcsec$
from the centre of the high-surface brightness elliptical galaxy M32. 
The Principal Investigators of the present archive data have elsewhere
presented an impressive colour-magnitude diagram of the field at $85\arcsec$.
Based on the same data we enlarge on their photometric analysis
and supplement with error estimators that more clearly show the
implications of severe image crowding on the stellar photometry.
We show that the faintest stars ($I \ga 25.0$, $V \ga 26.0$) are found too
bright by several tens of a magnitude.
For the field at $10\arcsec$ we conclude that
it is not possible to obtain reliable stellar photometry,
standard deviations being larger than 0.4~mag.
Artificial-star experiments show that only very few of the brightest stars
of the luminosity function can be expected to represent single objects,
the majority being either spurious or not as bright as measured.
The paper as such introduces and demonstrates
basic guide lines which may be used when dealing with stellar photometry of 
severely crowded fields.
\keywords{methods: data analysis -- galaxies: elliptical -- galaxies: individual: M32 -- galaxies: stellar content}
\end{abstract}
\section{Introduction}
\noindent
In order to increase our knowledge of extra-galactic stellar systems like
elliptical galaxies, we observe the average galaxy brightness,
which gives rise to the traditional broad-band colours 
and spectral-line indices.
In addition, we observe the variance of the galaxy brightness, i.e.,
the surface-brightness fluctuations (Tonry \& Schneider 1988).
Of course, the ultimative goal is to obtain
stellar luminosity functions and colour-magnitude diagrams,
i.e., accurate photometry of individual stars.

Until recently, colour-magnitude diagrams have been obtained solely
from ground-based imaging of stars in our local neighbourhood, i.e.,
field stars and stars in open- and globular clusters of our Galaxy,
in addition to approaches to
Local Group dwarf spheroidals and the Magellanic Clouds
({\it low surface-brightness} stellar systems)
from ground (e.g.\ Freedmann 1992) and space (e.g.\ Mighell \& Rich 1996).
Then Grillmair et~al.\ (1996) presented an impressive
colour-magnitude (CM) diagram of
M32, our closests elliptical and companion to the Andromeda galaxy,
based on Hubble Space Telescope imaging (archive data, ID 5233).
Using the same data we have been able to reproduce
the CM diagram of M32 which
in itself, however, was not the main goal of our work.
We have carried out an investigation of the reliability
of the CM diagram, that is, an estimation of the photometric
quality of the detected stars. The image crowding in dense fields
like that of the {\it high surface-brightness} elliptical M32
may cause spurious detections and severely reduce the reliability
of the photometry.
In the present paper we discuss specific aspects of the
photometric quality of extra-galactic stellar systems, 
aspects that supplement the discussion by Grillmair et~al.

In our latest paper (Sodemann \& Thomsen 1996)
we presented ground-based imaging of M32
obtained with the Nordic Optical Telescope
with a seeing of FWHM $\sim 1\farcs0$ using $0\farcs175$ pixels.
We identified a systematic variation in the $I$ and $B$-band
Surface-Brightness Fluctuations (SBF) of $0.2-0.3$~mag
in the radial range $ 10\arcsec \la R \la 50\arcsec$.
In our search for the stars responsible for this variation
we made a comparison of simulated stellar fields with
the ground-based SBF observations.
From this it became clear
that it was not possible to supplement the measurements of
the SBF with stellar photometry of the most luminous stars,
due to excessive crowding.

However, at least in some cases
the superb resolution of the Hubble Space Telescope's (HST) Wide Field
Planetary Camera 2 --
the Planetary Camera (PC) with seeing of FWHM $0\farcs1$ 
using $0\farcs05$ pixels --
makes it possible to resolve more than just the very brightest stars.
Grillmair et~al.\ (1996) presented a CM diagram based on HST-imaging
of a region $85\arcsec$ from the centre of M32.
From the surface brightness $\mu$ of M32 we can compare this
off-centre field with a central field at $10\arcsec$ from the centre.
Assuming that the same  type of stars are responsible for the 
surface brightness at the two distances,
the number of stars per area is $\sim 40$ times higher in the central field
($\mu^I_{85\arcsec}\simeq 19.8$ and $\mu^I_{10\arcsec}\simeq15.8$).
This merely indicates differing problems introduced by image crowding,
problems that may turn up 
when dealing with CM diagrams of fields at various distances from the galaxy
centre.  We shall return to this later.

When discussing luminosity functions and CM diagrams we are
interested in knowing to what magnitude limit we have actually 
been able to observe stars, a limit set mainly by a compromise between the 
image crowding and resolution, and the amount of exposure time. 
This magnitude limit is usually measured
by addition and retrieval of a small number of faint artificial stars,
that is, the test of (in)completeness.
However, whereas the traditional method of adding artificial stars 
estimates the detection probability it does
not measure the photometric quality of the detected stars or the effects
of spurious detections caused by severe image crowding.
That is, the estimate of the limiting magnitude
is based on information about the location
(the coordinates) of the added/recovered 
star alone and not on the magnitude of the recovered star.
For the data presented here, one important result of the image crowding is
the following: An artificial star of known magnitude
added to the programme frame may be recovered with a magnitude
up to 1$^{\rm m}$ brighter than its original magnitude, but
it turns out that it is not equally likely that this star will be
recovered with a magnitude 1$^{\rm m}$ fainter than its original magnitude.
(In Sec.~3 we show that the asymmetry in the distribution of recovered stars
cannot be explained as a result
of plotting the data as a function of magnitude).
Thus, the typical measure $\sigma$,
the standard deviation, of the accidental error at a given magnitude
is not an adequate description for this asymmetric distribution of recovered
stars and has to be supplemented with additional error estimates.

\begin{figure}[hbt]
\vspace{8.3cm}
\includegraphics{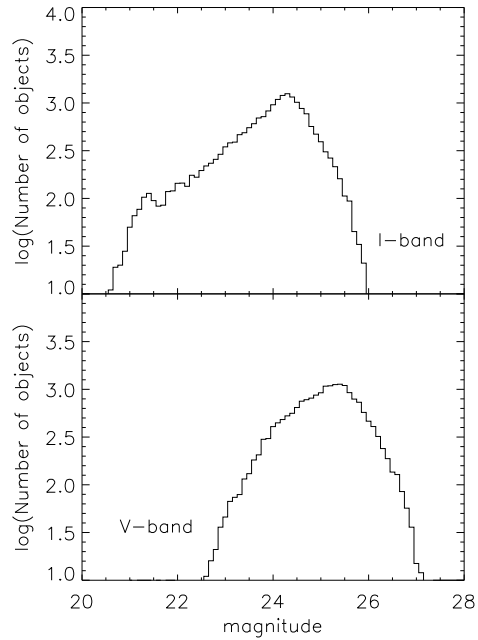}
\caption[s23gra]{\label{lumfu} Luminosity functions for objects identified
in I and V, respectively, in the HST PC-frame at $85\arcsec$ from the
centre of M32. A total of 19461 stars were found to match in I and V.
No correction for M31 has been made, see text.
The slope of the I-band luminosity function is found to be 0.42. (This value
is not used elsewhere in this paper).}
\end{figure}

\begin{figure}[hbt]
\vspace{13.0cm}
\includegraphics{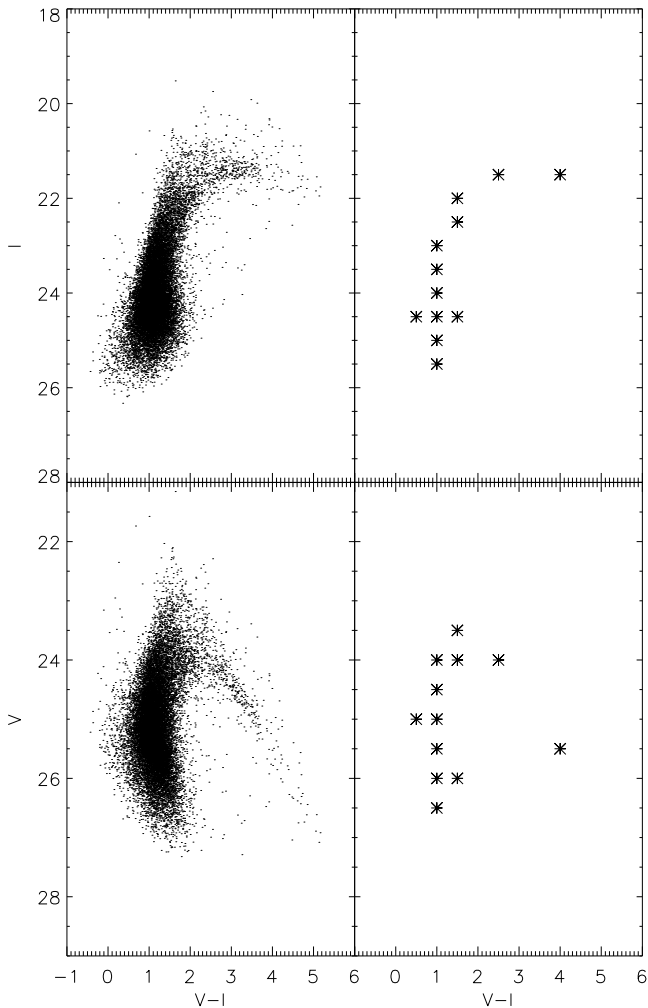}
\caption[s23gra]{\label{cm} Colour-magnitude diagram of the off-centre
field in M32. This figure was produced by matching the objects
in Fig.~\ref{lumfu}.
The 2$\times$12 
star signs show the magnitude and colour of
the stars in our artificial-star experiments.}
\end{figure}

Drukier et~al.\ (1988) raised the problem
of `bin jumping' and pointed out a way to account for this
when correcting a measured luminosity function for incompleteness.
LePoy et~al.\ (1993) also discuss the effects of severe image crowding,
which may produce an artificial enhancement of the bright end of the
measured luminosity function if not dealt with in an appropriate way
(see also Renzini 1992).

We shall concentrate on the implications of crowding on stellar photometry
of HST imaging
of extra-galactic systems like the compact elliptical galaxy M32.
We begin with a description of the data and image handling (Sec.~2).
After that follows a discussion on the artificial-star experiments in
connection with the off-centre field at $85\arcsec$ from the
centre of M32 (Sec.~3) and
the dense central field at $10\arcsec$ (Sec.~4).
We summarise the results in Sec.~5.
\section{Data and image handling}
\noindent
The observational basis for this paper is HST imaging
of an off-centre field at $85\arcsec$ from the centre
of M32 taken with the WFPC2 on October 22, 1994
(standard-calibration observations, archive data, ID 5233).
See Grillmair et~al.\ Fig.~1\&2 for an outline of the M32 PC field
on the sky.
The 2$\times$4 500~sec frames in F555W (V) and F814W (I), respectively,
were co-added through the cosmic-ray removing software of
Groth \& Shaya (1992) resulting in two programme frames.
An I-band programme frame of the field at $10\arcsec$ was produced in
a similar way from 3 exposures of 400~sec each (archive data, ID 5464).
Correction for vignetting was carried out according to 
Holtzman et~al.\ (1995a),
and standard magnitudes and colour were derived using
the calibrations in Holtzman et~al.\ (1995b).
We have adopted $A_I = 0.14$ for the galactic absorption 
(Burstein \& Heiles 1984, Cardelli et~al.\ 1989):
The luminosity functions and CM diagrams have been corrected for this.

The programme frames did not contain sufficiently bright, isolated stars
to define a proper point-spread function (PSF).
Therefore, model PSFs generated by the program Tiny Tim (Krist 1993) defined
the artificial stars added during the completeness tests
and the master PSF of the psf-fitting software by Stetson's (1987)
DAOPHOT~II package.

The completeness tests were carried out by adding stars 
distributed according to the observed CM diagram.
The 2$\times$12 star signs 
in Fig.~\ref{cm} for I and V, respectively,
outline the range of magnitudes and colours.
A total of 7110 artificial stars were added to each 
of the two programme frames. When carrying out an addition of stars,
only very few (in principle {\it one}, or several laid in a grid)
ought to be added at a time so as to avoid confusion between added stars and
an increase of the image crowding.
As an alternative, we split the addition of the artificial stars into two
and those stars
which happened to be added near another artificial star were removed.
The stars were added using our HST-Sim
routines, which scale and add the required number of PSFs
(the routines are implemented using Interactive Data Language, IDL).
The routines take into account the geometric distortion inherent in the
WFPC2 optical system, and
shifting and rebinning of the subsampled PSFs provide subpixel positioning.
Finally, Poissonian photon noise and Gaussian read noise are added.

The photometry of objects in the programme frames was obtained through
a procedure that included several passes of object identification
and measurement of magnitudes, using DAOPHOT~II.
First we worked through the identification of objects in the I-frame.
In order to identify as many objects as possible and
to ease the identification of faint objects we carried out
$\sim 6$ iterations, successively reaching fainter magnitudes.
For every two iterations we concatenated the lists of hitherto revealed stars,
subtracted those from the original frame, and continued from there.
The iterations ended when we reached a threshold of 4
standard deviations of the background level,
as calculated by DAOPHOT - at that point only a few hundred stars were 
identified and successfully subtracted by ALLSTAR and an additional 
iteration did not provide more objects.
The final ALLSTAR list of all stars in I was then applied
for star-subtraction in the V frame.
After having calculated flux-averaged positions based on the I and V list
of objects we again ran
ALLSTAR on both frames, now with fixed coordinates.
In this way a total of 19461 stars were found to match in V and I.
The photometry of the star-added frames was carried out in a way
identical to that of the programme frames.

Note, Grillmair et~al.\ ap\-pli\-ed Stet\-sons's (1994) pho\-to\-me\-try 
pack\-age ALL\-FRA\-ME.
They found that the global characteristics of the CM diagram
produced using DAOPHOT~II/\-ALLSTAR were essentially identical to those
found using ALLFRAME.

We have made no attempts to correct for the M31 background stars
as none of the conclusions of this paper depend on knowing this accurately.
According to Grillmair et~al.\ about 18\% of the $\sim 20000$ stars identified
in the programme frames originate in M31.
\section{Results from the off-centre field}
\noindent
The luminosity functions and CM diagrams of the off-centre field of M32 
are seen in Fig. \ref{lumfu} and \ref{cm}.
With the artificial-star experiments
as our basis we will now discuss the uncertainty
associated with the identified stars. 

\begin{figure}[hbt]
\vspace{10.0cm}
\includegraphics{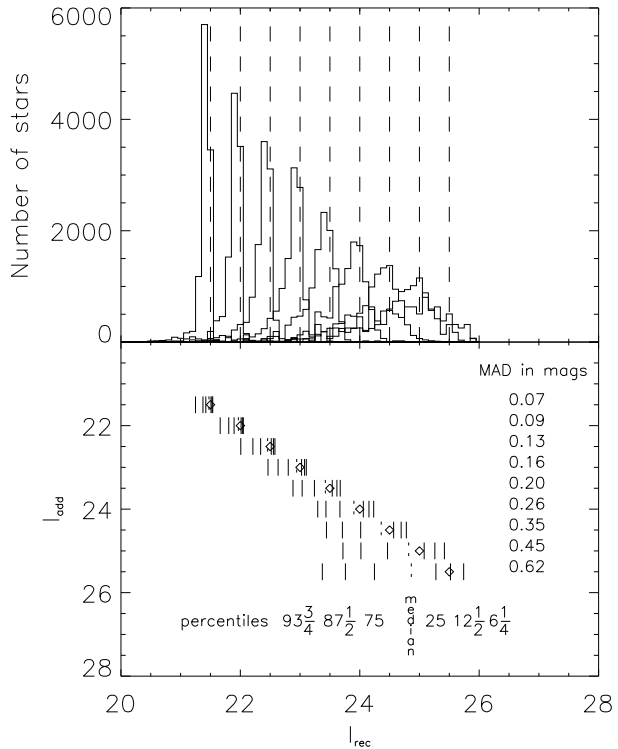}
\caption[s23gra]{\label{quarti} Results of our I-band artificial-star 
 experiments.
 For each magnitude group of added stars (vertical line)
 the histogram of recovered stars is
 shown. Note, the number-axis is arbitrary (Sec.~3). To clarify the
 implications of image crowding we plot several error indicators, in addition
 to the mean absolute deviation (MAD), of the magnitude of the recovered star
 $m_{\rm rec}$. 
 This demonstrates that
 the distribution of recovered stars is skewed towards brighter magnitudes,
 that is, the faintest objects are found too bright. This `bin jumping' was
 discussed by Drukier et~al.\ (1988).}
\end{figure}

\begin{figure}[hbt]
\vspace{10.0cm}
\includegraphics{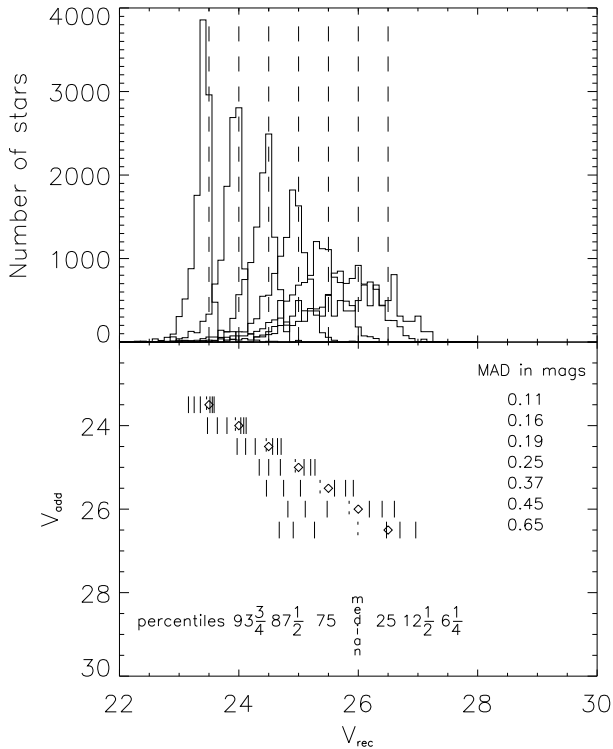}
\caption[s23gra]{\label{quartv} Same as Fig.~\ref{quarti} but now for the
 V-band.}
\end{figure}

In Fig.~\ref{quarti} and \ref{quartv} the magnitude of each group of 
added stars is indicated by a vertical line.
The magnitudes of the
recovered stars are binned into histograms of $0\fm1$ bins,
one histogram for each magnitude group of added stars.
Each of the histograms has been normalised
so that they all cover the same number of stars
(implying that the number-axis is arbitrary).
This illustrates that the brightest stars are easily identified,
whereas fainter stars are `spread out' from their original magnitude
(the vertical line) to surrounding magnitudes.
Firstly, note that for the faintest bins the probability of a star being found 
in a bin other than the one to which it was added is nonvanishing.
If we want to correct the luminosity function for incompleteness,
this `bin jumping' has to be taken into account (Drukier et~al.\ 1988).
Secondly, the `bin jumping' is clearly asymmetric with an extended tail
of bright measurements.
As the measurement errors are clearly not Gaussian
we shall replace the usual, but now inadequate,
standard deviation $\sigma$ by the mean absolute deviation (MAD).
(The MAD is
related to the median as the standard deviation is related to the mean,
but the MAD does not require that the distribution of the errors
is Gaussian and it is a more robust estimator than is the mean,
Press et~al.\ 1992).
In addition, we supplement with the lower and upper quartile,
and several percentiles, of the magnitude of the
recovered star $m_{\rm rec}$.
These numbers are indicated in Fig.~\ref{quarti} and \ref{quartv}
and illustrate that the faintest stars in the CM diagram
are very likely identified too bright by several tens of a magnitude
as compared to their true, unaffected magnitude.

Note, the straight lines
represented by diamond symbols in Fig.~\ref{quarti}, \ref{quartv},
and \ref{fig9} indicate the magnitude of
the added stars $m_{\rm add}$ and the expected magnitude of the
recovered stars {\it if they are not affected} by `binjumping',
i.e., $m_{\rm rec} = m_{\rm add}$.
In order to claim that we are able to carry out reliable photometry,
$m_{\rm add}$ must be a unique function of $m_{\rm rec}$,
and we must be certain that we are measuring the added star and not
a statistical lump of unresolved stars.

\begin{figure}[hbt]
\vspace{5.5cm}
\includegraphics{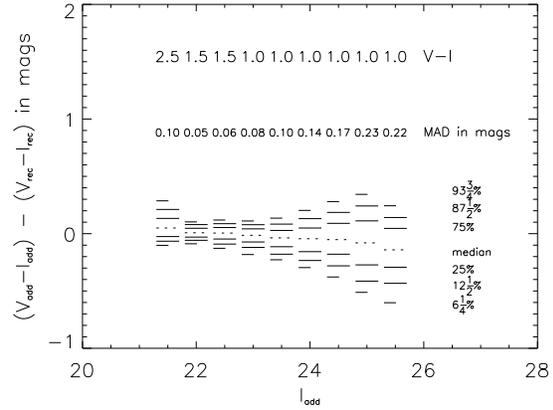}
\caption[s23gra]{\label{dVI} Colour deviation of recovered stars for
 the 9 artificial I-band magnitudes (error indicators are as in
 Fig.~\ref{quarti} and \ref{quartv}). 
 This shows that stars are very likely to be found with the correct colour. 
 The explanation is that
 the majority of stars in the CM diagram (Fig.~\ref{cm})
 have more or less identical colours $(V-I) \simeq 1.0$.}
\end{figure}

The distribution of the colour of recovered stars is, however, fairly symmetric
and much less wide than that of the two passbands individually
(Fig.~\ref{dVI}).
That is, the added stars which undergo `bin jumping' tend to keep their
original colour, mainly because they are most likely identified on top of a star
with a colour similar to their own -- the majority of stars in the CM
diagram are found along the giant branch with $(V-I) \simeq 1.0$.
We may therefore expect that the colours of the stars in the CM diagram
are correct.

\begin{figure}[hbt]
\vspace{5.0cm}
\includegraphics{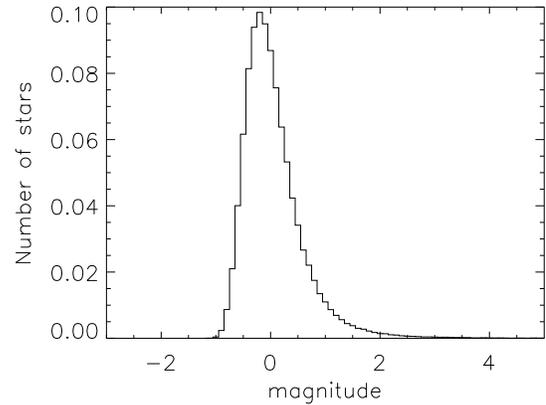}
\caption[s23gra]{\label{sym} Asymmetry that arises due to the effect of
 plotting a histogram of stars as a function of magnitude 
 $m = -2.5\log{f}$, where the distribution in flux $f$ is Gaussian
 (arbitrary number-axis).
 The asymmetry is opposite that of the histograms
 in Fig.~\ref{quarti} and \ref{quartv}.}
\end{figure}

Fig.~\ref{sym} illustrates that the asymmetric distribution in the histogram
of recovered stars is not a result of plotting the
number of stars as a function of magnitude.
We generated a Gaussian distribution of fluxes according to
${\rm d}N = N_0{\rm e}^{-(f-f_0)^2/2\sigma^2}{\rm d}f$,
where $N_0 = 10^6$, $f_0 = 1.0$, and $\sigma = 0.4$.
The corresponding magnitudes $m = -2.5\log{f}$ were binned into
$0\fm1$ bins.
The increase in the number of stars fainter than $f_0$ ($m > 0$)
exceeds a small increase just below $m = 0$ and is clearly opposite 
the asymmetry seen in Fig.~\ref{quarti} and \ref{quartv}.
Note that Secker \& Harris (1993) adopt a Gaussian function to represent
the measurement error as a function of magnitude.

The limiting magnitudes at half probability are
$V_{1/2} = 26.2$ and $I_{1/2} = 25.1$ for the M32 off-centre field.
The limiting magnitudes are based on {\it all} the
stars plotted in Fig.~\ref{quarti} and \ref{quartv}, respectively,
i.e., regardless of the associated error estimate.
Note, the standard deviation $\sigma$, as calculated by ALLSTAR,
amounts to only $0\fm12$ and $0\fm16$ at the limiting magnitudes
for I and V, respectively, 
whereas the MAD is higher by a factor of 3.8 and 2.8 as compared
to the standard deviation (for a Gaussian distribution 
${\rm MAD} \simeq 0.8 \sigma$).
This merely indicates that in the present investigation 
the standard deviation, as given by ALLSTAR,
may be considered too optimistic.
\section{The M32 centre}
\begin{figure}[hbt]
\vspace{6.0cm}
\includegraphics{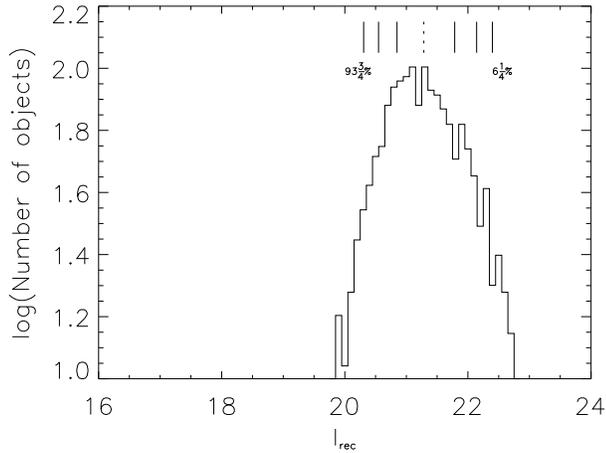}
\caption[s23gra]{\label{histo_i} Luminosity function for objects
 identified in I in the HST PC-frame centered on M32.
 A total of 1606 objects were found in an annulus of area $80\arcsec^2$ 
 and radius $\sim 10\arcsec$. The median indicated at the top is $21\fm3$.
 Only a few of the very brightest stars can be considered to be 
 single objects.}
\end{figure}

\noindent
Based on the investigation of the off-centre field
it is possible to predict
the number of identified stars in a central field at about $10\arcsec$,
where the density is $\sim 40$ times higher than in the off-centre field
(Sec.~1).
The I-band luminosity 
function of the off-centre field starts to turn over
at a number density of $\sim 1000$ 
per $0\fm1$ in the PC frame (Fig.~\ref{cm}).
In a field with density 40 times higher than this, 
the turn-over would correspond to a
magnitude of $\sim20\fm5$, or at the very brightest end of
the luminosity function where the number density drops to zero.
An interesting question is whether it is possible to find 
stars of $20\fm4$ in the central region of M32.
(This magnitude corresponds to $M_I \simeq -4.0$, 
the expected tip of the red-giant branch
for metal-poor stars, e.g.\ Lee et~al.\ 1993, and could provide an
estimate of the distance to the galaxy).
To address this question we have investigated a central field in M32
(archive data, ID 5464), 
the analysis being virtually identical to that of the off-centre field.
(The addition of artificial stars was split up so that no more than one
magnitude group at a time, corresponding to $\sim 700$ stars, was added to
the programme frame).
The luminosity function of the central field is seen in Fig.~\ref{histo_i}.

\begin{figure}[hbt]
\vspace{12.0cm}
\includegraphics{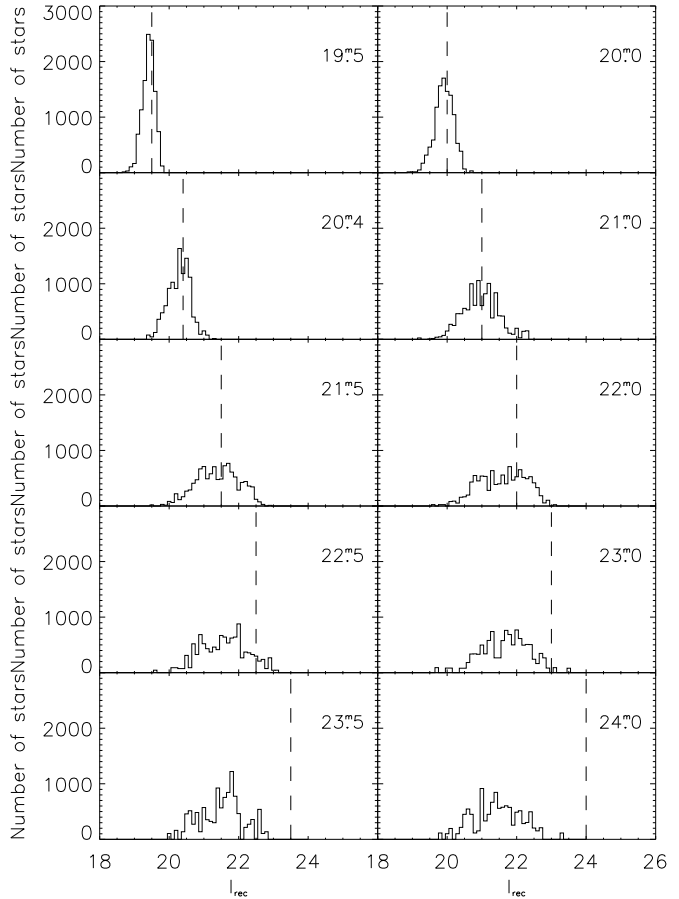}
\caption[s23gra]{\label{diagr}Results of our I-band artificial-star
 experiments, now for
 the central region of M32, approximately $10\arcsec$ from the centre.
 Again, the number-axis is arbitrary (Sec.~3).}
\end{figure}

Apparently we do see stars of magnitude $20\fm4$ (Fig.~\ref{histo_i}). However,
when we discuss the central field, we also have to consider the confusion
that spurious detections due to SBF may cause, i.e., the confusion between
single objects and detection of statistical lumps in the background of
unresolved stars
(they both have the shape of the PSF).
This problem is illustrated in Fig.~\ref{diagr} 
where we see that bright objects around, e.g., $20\fm4$ are indeed 
detected and 
located in the recovery tests with artificial stars of magnitude $21\fm0$ and
fainter.
The completeness tests show that {\it if} stars of 
magnitude $20\fm4$ are present
in the centre of M32, then we will be able to identify them 
(the data in Fig.~\ref{diagr} imply a limiting magnitude at half probability
of $I_{1/2} = 21.6$ for the central field of M32).
What we do not know is how large a fraction of the 
supposedly identified bright stars in Fig.~\ref{histo_i}
are merely compounds of fainter, unresolved stars.
The limiting magnitude itself gives us no information about that.

\begin{figure}[hbt]
\vspace{6.0cm}
\includegraphics{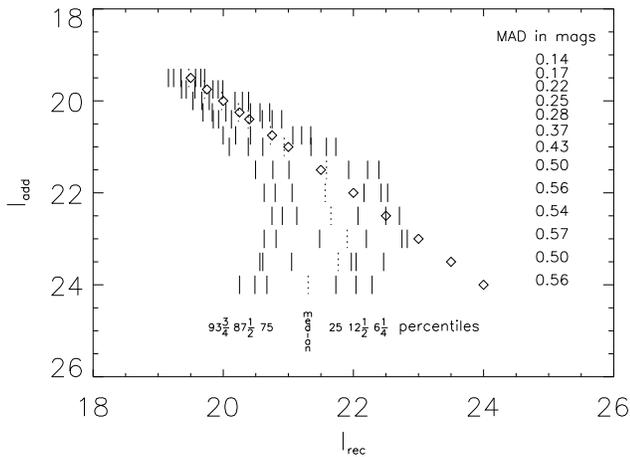}
\caption[s23gra]{\label{fig9} Same data as Fig.~\ref{diagr}
 (besides three extra sets of artificial stars)
 but plotted in a slightly different way
 (error indicators are as in Fig.~\ref{quarti} and \ref{quartv}).
 As emphasized in the text,
 $m_{\rm add}$ must be a unique function of $m_{\rm rec}$,
 and we must be certain that we are measuring the added star and not
 a statistical lump of unresolved stars, before we can claim that we
 are able to carry out reliable photometry.
 This is clearly not the case for
 magnitudes fainter than $\sim 20\fm5$. From that magnitude and onwards 
 it is no longer
 possible to distinguish between the distributions of recovered stars.
 Therefore, only a few of the very brightest stars 
 with $I \la 20.5$ in Fig.~\ref{histo_i}
 may be expected to be single objects.}
\end{figure}

As will be evident from the following, we can only expect very few of the
brightest stars of the observed luminosity function to be single objects.
Fig.~\ref{fig9} illustrates that it is not possible to distinguish
between the distributions of stars added at $\sim 20\fm5$ and fainter.
That is, when we identify a star with magnitude
$\sim 20\fm5$ or fainter, then we can by no means tell the true 
magnitude of that star.
The completeness tests imply a limiting magnitude of
$I_{1/2}=21.6$ as mentioned above, but this limit only 
indicates the detection probability.
We are inclined to disregard the stars in Fig.~\ref{histo_i}
 that are fainter than
$\sim 20\fm5$ (i.e., even stars brighter than $I_{1/2}$),
since we have no reason to believe that they are single point sources.

\begin{figure}[hbt]
\vspace{6.0cm}
\includegraphics{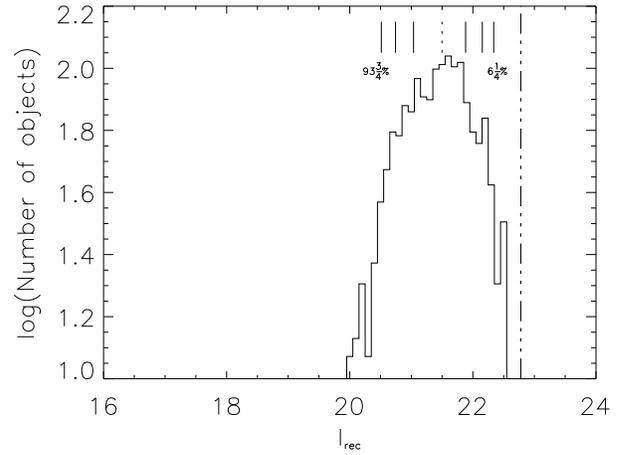}
\caption[s23gra]{\label{histo_sim} Luminosity function of a simulated
 stellar field with $\mu^I_{10\arcsec}\simeq15.8$, as observed for the
 central field, but generated from only {\it one} type
 of stars, namely stars of magnitude
 $\bar m_I = 22.78$ (vertical line), see text.
 Note the similarities to
 the observed luminosity function (Fig.~\ref{histo_i}).
 The median indicated at the top is $21\fm5$.
 (The histogram above has been normalised to the same number of objects as 
 that of Fig.~\ref{histo_i}).}
\end{figure}

This is further emphasised by a simple simulation of the stellar field at
$10\arcsec$:
As the stellar content we adopt only one type of stars, namely stars
of magnitude $\bar m_I = 22.78$ (Sodemann \& Thomsen 1996).
Our HST-Sim routines add the required number of such `fluctuation stars'
(PSFs) of magnitude $\bar m_I$ to reach the average surface brightness of
$\mu^I_{10\arcsec}\simeq15.8$ as observed for the central frame,
and photometry is carried out by DAOPHOT~II.
The luminosity function of this simulated stellar field is seen
in Fig.~\ref{histo_sim} and should be compared with the
luminosity function of the observed central field (Fig.~\ref{histo_i}).
The similarities are striking and tell us that the objects in
Fig.~\ref{histo_i} may all, except for a very few of the brightest,
be described as mere fluctuations and
not as single point sources.
Based on Fig.~\ref{histo_sim} we may explain the observed
`skewness' in Fig.~\ref{fig9} as follows.
Those of the faint added stars which were located
on top of a bright fluctuation
have inevitably been measured too bright and therefore end up in the left part
of the `magnitude bands' of recovered stars
(a magnitude band being defined by the 93.75 and 6.25 percentiles), whereas
the stars in the right part of the `magnitude bands'
were located on top of a less bright fluctuation.

We do not correct the observed luminosity functions for incompleteness and
binjumping for the following reasons.
Concentrating on the $I$-band,
for the field at $85\arcsec$ the histograms of recovered stars are
distinguishable for magnitudes that correspond to the giant branch and
down to $I \simeq 24.0$ where incompleteness starts to set in
(Fig.~\ref{lumfu} and \ref{quarti}).
In this paper we do not require the incompleteness-corrected luminosity
function, and correction for binjumping would only affect the
magnitudes around the limiting magnitude.
For the field at $10\arcsec$, the luminosity function is
comparable to any of the distributions of added test stars fainter than 
$\sim 20\fm5$,
and an attempt to deconvolve the effects of `binjumping'
would thus be an ill-posed problem.

\section{Conclusion}
\noindent
In this paper we have discussed the effects that severe image crowding
has on stellar photometry.
The investigation covered two fields of the high-surface brightness
elliptical galaxy M32 observed with the HST's PC
(archive data, ID 5233 and 5464).

We have carried out a large number of artificial-star experiments,
that is, addition and retrieval of artificial stars.
Our experiments clearly show the presence of
`binjumping': The faint stars added to the programme frames
are recovered too bright; the fainter the
star and the denser the field, the bigger is the effect of `binjumping'.
The traditional indicator of (in)completeness of the
photometry, $m_{1/2}$, merely provides a detection probability but not
the quality of the photometry.

For the less dense field at $85\arcsec$ from the centre of M32, the effects
of image crowding show up around the limiting magnitude $I_{1/2}$
($25\fm1$).
Stars fainter than that are found too bright by several tens of a magnitude
as compared to their true magnitude,
whereas their colour is mainly unaffected.
For the far more dense field at $10\arcsec$ from the centre,
`binjumping' affects stars that are even brighter than $I_{1/2}$ ($21\fm6$).
As our Fig.~\ref{fig9} demonstrates, only the very brightest stars with
$I \la 20\fm5$ may be considered to represent single objects.
Thus, we do measure single
stars of, e.g., $20\fm4$ (corresponding to $M_I \simeq -4.0$, 
the expected tip of the red-giant branch
for metal-poor stars) in the off-centre field at $85\arcsec$,
but based on the present data 
it is not possible to claim that for the central field at $10\arcsec$,
where the density is 40 times higher.

Finally, we would like to emphasize that the data analysis, which has been
introduced and described in the present paper in relation to HST imaging,
should be carried out for essentially all imaging of severely crowded fields.
\begin{acknowledgements}
We wish to thank P.\ B.\ Stetson for the benefits of 
using his photometry package DAOPHOT~II.
\end{acknowledgements}

\end{document}